\documentclass[fleqn,twoside]{article}
\usepackage{espcrc2}

\usepackage{graphicx}
\usepackage[figuresright]{rotating}

\newcommand{\AmS}{{\protect\the\textfont2
  A\kern-.1667em\lower.5ex\hbox{M}\kern-.125emS}}

\newcommand{\nuebar}{\overline{\nu}_{e}}

\def\adm2{\Delta{{m}^2_{{\rm atm}}}}\def\sdm2{\Delta{{m}^2_{{\rm sol}}}}
\def\be{\begin{equation}}
\def\ee{\end{equation}}
\def\Dm2{\Delta{m}^2}
\def\t13{\theta_{{13}}}

\def\simgt{\lower.5ex\hbox{$\; \buildrel > \over \sim \;$}}
\title{Chasing $\theta_{13}$ with  new reactor neutrino experiments}

\author{Th. Lasserre\address[MCSD]{DSM/DAPNIA/SPP, CEA/Saclay, 91191 Gif-sur-Yvette, France}}%
       
\begin{document}

\begin{abstract}
%
It is now widely accepted that a new middle baseline disappearance reactor neutrino
experiment with multiple detectors could provide a clean measurement of the $\theta_{13}$ mixing angle, 
free from any parameter degeneracies and correlations induced by matter effect and the unknown leptonic Dirac CP phase. 
The current best constraint on the third mixing angle comes from the Chooz reactor neutrino experiment
$\sin^{2}(2\theta_{13})<0.2$ (90$~\%$~C.L., $\Delta m_{\rm atm}^{2}=2.0 \, 10^{-3}$ eV$^{2}$).
Several projects of experiment, with different timescales, have been proposed over the last two years all around the world. 
Their sensitivities range from $\sin^{2}(2\theta_{13})<$ $0.01$ to $0.03$, having thus an excellent discovery potential 
of the $\nu_e$ fraction of $\nu_3$.
\vspace{1pc}
\end{abstract}

\maketitle
\section{Physics motivations}
\par Considering only the three known families, the neutrino mixing matrix is parametrized by the three mixing angles 
($\theta_{12}$, $\theta_{23}$, $\theta_{13}$). $\theta_{12}$ has been measured to be large ($\sin^{2}(2\theta_{12})\sim 0.8$) 
by the combination of the solar neutrino experiments and the KamLAND long baseline reactor neutrino experiment \cite{Eguchi:2002dm}.
$\theta_{23}$ has been measured to be close to maximum ($\sin^{2}(2\theta_{23})>0.9$) by atmospheric neutrino experiments 
as well as the long baseline accelerator neutrino experiment K2K~\cite{K2K}.
The mixing angle $\theta_{13}$ coupling both solar and atmospheric sectors has only been upper constrained, mainly by the 
Chooz \cite{choozlast} reactor neutrino experiment ($\sin^{2}(2\theta_{13})<0.2)$. 

\par On the one hand the large size of both $\theta_{12}$ and $\theta_{23}$ 
indicates a strong difference between leptonic and quark mixings. 
On the other hand the smallness of $\theta_{13}$ testifies the peculiarity of the neutrino sector.
Though the value of $\theta_{13}$ is of fundamental interest to understand the leptonic mixing, it is also very important to 
define the future experimental program. Indeed all CP-violating effects that could be measured 
by neutrino oscillation experiments are proportional to $\sin^{2}\theta_{13}$. The knowledge of the order of magnitude of 
$\theta_{13}$ is thus important for the design of the future experiments dedicated to CP violation.
\section{Reactor neutrinos}
\par Nuclear reactors produce only $\nuebar$ through beta decays of the fragments of  the fissionable materials.
For instance, the light water reactors (BWR and PWR) fuel mainly consists of $^{235}$U and $^{239}$Pu, 
which undergo thermal neutron fission. The dominant natural uranium isotope, $^{238}$U, is fissile only for
fast neutrons (threshold of 0.8 MeV) but it also generates fissile $^{239}$Pu by thermal neutron capture. 
The  $^{241}$Pu isotope  is produced in a manner similar to $^{239}$Pu. Fuel composition evolves with time (burn-up).
\par Reactor antineutrinos are detected through the inverse neutron decay 
$\bar\nu_e + p \rightarrow e^+ + n~$ (threshold of 1.806~MeV), with a cross section
$\sigma(E_{\rm e^+})\simeq (2\pi^2\hbar^3) / (m_e^5 f \tau_n)p_{\rm e^+}E_{\rm e^+}$,
where $p_{\rm e^+}$ and $E_{\rm e^+}$ are the momentum and the energy of the
positron, $\tau_n$ is the  neutron lifetime and $f$ is the neutron decay phase space factor.
For a typical PWR averaged fuel composition $^{235}$U (55.6~\%), $^{239}$Pu (32.6~\%), $^{238}$U  (7.1~\%) 
and  $^{241}$Pu (4.7~\%), and a thermal power $P_{\rm th}$ (GW$_{\rm th}$),  
the number of fissions per second $N_{\rm f}$ is given by 
$N_f = 3.06 \cdot 10^{19} {\rm s}^{-1} \, P_{\rm th}[{\rm GW}]$. 
The energy weighted cross section amounts to
$<\sigma> _{\rm fis} = 5.825 \cdot 10^{-43}~{\rm cm}^2 \,  {\rm per} \, {\rm fission}$.
The event rate at a distance $L$ from the source, assuming no oscillations, is given by 
$R_{\rm L}=N_{\rm f} <\sigma>_{\rm fis} n_{\rm p} \cdot~1/(4\pi L^2)$,
where $n_p$ is the number of (free) protons in the target.

\par Experimentally one detects the very clear signature of the coincidence signal of the prompt positron followed in 
space and time by the delayed neutron capture.  This allows to strongly reject the accidental backgrounds.  
The visible energy seen in the detector is given by  $E_{\rm vis}=E_{\rm e^+} + 511$~keV.

\par Reactor neutrino experiments measure the survival probability 
$P_{\nuebar \rightarrow \nuebar}$ of the $\nuebar$  emitted by nuclear power stations.
The $\nuebar$ disappearance probability is not sensitive to the Dirac CP phase $\delta$ (see Eq.~\ref{3nuSP}). 
Furthermore, thanks to the low neutrino energy (a few MeV) as well as the short baseline considered (a few~km), 
matter effects are negligible \cite{minakatareactor2002,Lindnerreactor}.  In the case of normal hierarchy, 
the $\nuebar$ survival probability  can be written 
\begin{eqnarray}
\label{3nuSP}
P_{\nuebar\to\nuebar}  = 1-4\sin^2\theta_{13}\cos^2\theta_{13}\sin^2 \frac{\Delta{m}^2_{31}L}{4E} & \nonumber \\
-  \cos^4\theta_{13}\sin^2(2\theta_{12})\sin^2\frac{\Delta{m}^2_{21}L}{4E}  & \nonumber \\ 
+2\sin^2\theta_{13}\cos^2\theta_{13}\sin^2\theta_{12}  & \nonumber \\ 
   \left(\cos \frac{(\Delta{m}^2_{31}-\Delta{m}^2_{21})L}{2E}-\cos \frac{\Delta{m}^2_{31}L}{2E} \right) ,  & 
\end{eqnarray} 
where the first two terms in Eq.~\ref{3nuSP} contain respectively the atmospheric ($\Dm2_{31} = \adm2$) 
and solar driven ($\Dm2_{21} = \sdm2$) oscillations. The third term is an interference 
between solar and atmospheric contributions, which has a detectable influence 
only in a small region of the space of mass and mixing parameters \cite{HLMA}. 
For the considered case the first term dominates, which leads to 
a pure $\theta_{13}$ measurement.

\subsection{The new experimental concept}
\par In order to improve the Chooz constraint on $\theta_{13}$ \cite{choozlast}, 
one has to use a set of at least two identical detectors. 
The first one is at a few hundred meters from the reactor cores. It monitors the $\nuebar$ flux without oscillations. 
The second one is between 1 and 2~km.  It searches for a departure from the overall $1/L^2$ behavior of the $\nuebar$ 
energy spectrum,  induced by oscillations. At Chooz, the reactor induced systematic error amounted $1.9~\%$ of the  $2.7~\%$ 
total experimental uncertainty. This error cancels with the multiple detectors concept.
Furthermore, additional experimental errors are being either reduced or cancelled when using identical detectors. 
Thus, with slight technical innovations with respect to previous solar and reactor experiments,  
an overall systematic error of $0.6~\%$ (no background) is achievable \cite{DoubleChoozLOI,DoubleChoozLOIUS}. 
In order to fully utilize the potential of such detectors, the statistical error has to be 
decreased to a similar amount with respect to Chooz, providing 50,000 neutrinos events in the far detector, 
and millions in the near one. The close detector could then be used not only for fundamental research, 
but also for some investigations of the International Atomic Energy Agency (IAEA) \cite{DoubleChoozLOI,Bernstein}.  

\par The signature for a $\nuebar$ event may be mimicked by background events
which can be divided into two classes: accidental and correlated events.
The former can be reduced by a careful selection of the materials
used to build the detector. In addition this background is measurable
in-situ, and its subtractions lead to a small systemactic error. 
Two processes mainly contribute to the correlated background: 
$\beta$-neutron cascades and very
fast external neutrons. Both types of events are coming from spallation
processes of high energy muons. 
The correlated background can be reduced by locating the detector deep underground. 
It could also  be measured during the reactor ``OFF'' periods (this concerns the 
single and more weakly the double core power stations).
\subsection{Complementarity with Superbeams}
Forthcoming accelerator neutrino experiments, or Superbeams, will search for a $\nu_{\rm e}$  appearance signal.  
The observation of a $\nu_{\rm e}$ excess in an almost pure $\nu_{\rm \mu}$ neutrino beam at any accelerator experiment would 
be major evidence for a non-vanishing $\t13$.   But unfortunately, in addition to the statistical and systematic 
uncertainties, correlations and degeneracies between  
$\theta_{13}$, $\theta_{12}$,  sgn($\Dm2 _{31}$), and the CP-$\delta$
phase degrade the accessible knowledge on $\theta_{13}$~\cite{minakatareactor2002,Lindnerreactor}. 

Even though appearance experiments seem to be the easiest way to measure
very small mixing angles, as might be the case for $\theta_{13}$, it is of great interest 
to get additional information with another experimental method, 
and different systematic uncertainties. 
To accomplish this goal, both reactor and accelerator programs should
provide the required independent and complementary results~\cite{WhitePaper}.
\section{The current proposals}
\par Since the first KamLAND results, a community of physicists is preparing a new experimental program. 
A few international workshops have been organized in Asia, Europe, and US, leading to the publication 
of a White Paper \cite{WhitePaper}. Several sites have originally been proposed, from west to east : Diablo Canyon and
Braidwood (US), Angra (Brazil), Chooz, Cruas, and Penly (France), Krasnoyarsk (Russia), Daya Bay (China), and 
Kashiwazaki (Japan). I present below a personal selection of the current projects. 
\subsection{Kr2Det (Russia)}
\par The pioneering Kr2Det aimed to use two liquid scintillator detectors, located at 100~m and 1,000~m from the Krasnoyarsk 
underground reactor \cite{Kr2det}. This site takes advantage of the two existing cavities 
protected by an overburden of 600~mwe,   as well the 50/7 days ON/OFF cycle 
allowing in situ background measurements. 
However, due to the low thermal power of the $1.6~$GW$_{\rm th}$ single reactor,  two large 50 ton detectors are required. 
A sensitivity of  $\sin^{2}(2\theta_{13})<0.015$ (90~\%~C.L.,  $0.5~\%$ systematic errors, 3~years of operation) has been 
estimated. Unfortunately, political issues prevent any international effort at this site.
\subsection{Double-Chooz (France)}
\par The experiment site is located close to the twin reactor cores 
of the Chooz-B nuclear power station (PWR, 2$\times$4.3$~$GW$_{\rm th}$), operated by the French  
company Electricit\'e de France (EDF). 
The two almost identical detectors contain a 10 ton fiducial volume of liquid scintillator doped
with 0.1~\% of Gadolinium (Gd). The underground laboratory of the first Chooz experiment,
located 1.05~km (overburden of 300~mwe) from the cores is going to be used again. This is the main
 advantage of this site. The second detector will be installed
close to the nuclear cores. An artificial hill of about 20~m height has to be erected.
 (At 150~m the required overburden is 60~mwe.)
\par The detector design is an evolution of the Chooz detector \cite{choozlast}.
In order to increase the exposure to 60,000 events at Chooz-far it is planned
to use a target cylinder 2.3 times as big as in Chooz.
The near and far detectors will be identical inside the PMT supporting
structure. Starting from the center of the target the detector elements
are as follows: 
The neutrino target: A 120~cm radius, 280~cm height, 8~mm thick acrylic
cylinder, filled with 0.1~\% Gd loaded liquid scintillator.
The baseline of the scintillator being developed 
is a mixture of 20~\% of PXE and 80~\% of dodecane, with small quantities
of PPO and bis-MSB added as fluors. 
The $\gamma$-catcher: A 60~cm buffer of unloaded liquid scintillator, 
with the same light yield, enclosed 
in a 10~mm thick acrylic cylinder. 
The role of this {\it new} region is to get the full positron
energy, as well as most of the neutron energy released after neutron
capture. 
The non scintillating buffer: A 95~cm buffer of non scintillating
oil, in order to decrease the accidental backgrounds (mainly from PMTs radioactivity) 
and the structure supporting the 512 PMTs.
The muon veto: A 60~cm veto region filled with liquid scintillator
for the far detector, and a  larger one (1 to 2~m) for
the near detector.
The external shielding: 15~cm of steel surrounding the far
detector, and a 1~m low radioactive sand or water layer for
the near detector. 
An outer veto gas proportional chambers)  surrounding the cuve 
 will be added at the near detector site to tag cosmic ray induced backgrounds.

\par The dominant error is the relative normalization between the
two detectors, that  is expected to be less than 0.6~\%. The main contributions
come from the solid angle (0.2~\%), the volume (0.2~\%), the density and
H/C ratio (0.15~\%), the neutron detection efficiency and energy measurement
(0.2~\% and 0.1~\%), the e$^{+}$-n time delay (0.1~\%) and
the dead time (0.25~\%).

\par A Monte-Carlo study shows that the correlated events are 
the most severe background source. 
In total the background rates (accidental+correlated) for
the near detector would be between 9/d and 23/d, for 60~mwe
 overburden. For the far detector the total background is estimated between 1/d and 2/d. 
However, the accidental part, which dominates,  can be measured and subtracted, 
leading to a small systematic error.
This can be compared with
the signal rates of $4,000$/d and $80$/d in the near and far detectors.  
\par The sensitivity is $\sin^{2} (2\theta_{13}) < 0.025$ (90$~\%$~C.L., for $\Delta m_{\rm atm}^{2}=2.4 \, 10^{-3}$ eV$^{2}$, 
3~years of operation) in the no-oscillation case. The discovery potential is around 0.04 (3-$\sigma$). 
If $\sin^{2} (2\theta_{13})$=0.1 a rate only analysis could reject the no-oscillation scenario at 2.6$\sigma$, whereas 
a shape+rate analysis could reject the no-oscillation scenario at about 6$\sigma$.  

\par Double-Chooz is a collaboration between France, Germany, Italy, Russia and US. 
A Letter of Intent has been released in may 2004 \cite{DoubleChoozLOI}, 
as well as a proposal for the US participation \cite{DoubleChoozLOIUS}.  
The experiment has been approved in France. Double-Chooz has started its design phase. 
The plan is  to start taking data at the Chooz-far in 2007 (no civil construction), 
and at Chooz-near in 2008. 
\subsection{Kaska (Japan)}
\par The Kaska experiment~\cite{Kaska} could be located close to the Kashiwazaki,-Kariwa nuclear power station (BWR, 24.3 GW$_{\rm th}$), 
 which is the most powerful in the world. The plant is composed of 7 cores divided in 2 clusters spread by 2~km. Thus, 
two near detectors are mandatory (each at 400 m from a cluster). Two options are being considered for the far detector location : 
 1.3~km or 1.8~km.  
In that case, being at the optimum baseline of 1.8~km could be a disadvantage since the natural size of the 
detector is strongly constrained by the shaft hole diameter.
Since no natural hill exists around the power station, three 6~m diameter shaft holes have to be excavated and equipped. 
The required depth is 70~m for both near shafts,  and 200~m (250)  for the far one at 1.3~km (1.8). 
\par The Kaska design is similar to the Double-Chooz one: a 8 ton target of Gd doped liquid scintillator and a 
$\gamma$-catcher region enclosed in a double acrylic cylinder, a gamma shielding, a PMT supporting structure, 
and a weak scintillating region acting as a muon veto. The external layer is an iron shielding protecting 
from the external radioactivity. 

\par The systematic error foreseen is between 0.5 and 1~\%. 
A statistical error of $0.4~\%$ ($0.6~\%$) is expected at 1.3~km (1.8) after three years of operation. 
The corresponding sensitivity is expected to be between 
$\sin^{2} (2\theta_{13})< 0.017-0.027$ ($90~\%$~C.L., depending on the true value 
of $\Delta m_{\rm atm}^{2}$), in the no-oscillation case.

\par Kaska is a Japanese collaboration. R\&D and geological studies are ongoing until 2005. If approved, the plan is to 
construct the detectors in 2006-08 and to start data taking end of  2008.  
\subsection{DaYa-Bay (China)}
\par The DaYa-Bay experiment \cite{Dayabay} could be located in the Guang-Dong Province, 
close to the DaYa Bay nuclear power plant 
(PWR, two twin cores of $2.9$ GW$_{\rm th}$, for a total of $11.6$ GW$_{\rm th}$).
A third twin unit is planned to be in operation in 2010. 
Three identical liquid scintillator detectors will be used, two ``near'' and one ``far'' detectors.
Since the power station has been built near a mountain, the plan is to excavate an horizontal  
tunnel and several experiment halls. 
Those civil constructions would provide an excellent overburden for the detectors: 
 $400$~mwe at a distance of $300$~m from the cores, and  $1,200$~mwe at a distance of about 
$1,500-2,000$~m. This provides flexibility to optimize the detector rooms location 
according to the volatile  best fitted $\Delta m_{\rm atm}^{2}$ value.
A detector unit contains three main regions: a 8 tons target 
in a transparent Plexiglas container filled with a 0.09$~\%$ Gd-loaded scintillator; 
a second intermediate region equipped with hundreds eight-inch PMTs, used to protect 
the target from PMT radioactivity as well as to contain the gamma rays from neutron capture; 
a third region optically separated acting as a cosmic-ray muon veto shielding, equipped eight-inch PMTs. 
To protect the detector from natural radioactivity of the rock, the steel vessel could be
surrounded by 1 m of low radioactivity sand and covered by 15~cm of iron.  Thus, the background 
rates could be  suppressed down to $0.2-0.3$ events per day per ton.
The particularity of this experiment is to use the so-called ``movable detector'' concept. 
The three detectors could be swapped for cross calibration. The question of the systematic error introduced 
when moving such sensitive machines remains open, however.
\par Several schemes of the detector locations have been studied, but resulting to similar sensitivities.  
For three years of operation, and according to this preliminary configuration \cite{Dayabay}, 
the 3-$\sigma$ discovery  potential is expected to be 
$\sin^{2}(2\theta_{13})\sim0.03$ ($\Delta m_{\rm atm}^{2}=2.5 \, 10^{-3}$ eV$^{2}$). 
Geological studies as well as safety investigations are currently being done; R\&D  proceeds until 2005. 
The plan is to start construction in 2006, in order to operate a first near detector in 2008.
\subsection{Braidwood (US)}
The Braidwood experiment \cite{Braidwood} would be located close to the Braidwood twin nuclear station 
(BWR, 2$\times$3.6$~$GW$_{\rm th}$), in Illinois. The area surrounding the power plant
has a very flat topology, thus two 10~m diameter shafts and two 12-32~m  detector rooms
have to be excavated. The two laboratories could have an overburden of 
450~mwe (120~m underground), providing the same background contamination in each detector. 
The ground, composed of dolomite limestone, is convenient for excavation. 
The detector rooms can thus be optimized according to the future knowledge of $\Delta m_{\rm atm}^{2}$. 
The plan is to have $\geq$1 near detector of 25-50 tons (fiducial mass) at 270~m in the near shaft, 
and $\geq$2 far detectors identical to the near one,  at $\sim$1.8~km in the far shaft.  
The detectors could be constructed above ground, and moved to the experimental sites on a platform transporter.
Then  a 750-ton capacity crane would carry the detectors underground. 
This operation could be repeated several times in the 
lifetime of the experiment in order to swap the detectors for cross calibration.
The detector design could be different to Double-Chooz and KASKA by omitting the $\gamma$-catcher region. 
\par The sensitivity depends on the final layout of the experiment which is still uncertain. 
The goal is to reach $\sin^{2}(2\theta_{13})<0.01$ ($90~\%$~C.L.) if no-oscillation is assumed.  
Currently, geological studies are being done. If approved, 40 months would be needed for the simultaneous 
construction of the detectors and the underground laboratories. 
\subsection{Angra (Brazil)}
\par A second generation of experiments, with a large target mass, and extremely low and very well understood backgrounds, 
might be needed in the future: i) if a first generation experiment observes a signal close to its sensitivity limit; 
ii) in complement to the Superbeam program in order to start to probe the ($\theta_{13}$,$\delta$-CP) plane 
\cite{Deltareactor}. 
This is an experimental challenge that could only be addressed  after the generation of experiments presented in the previous sections. 
\par The Angra experiment \cite{Angra}, near the 6 GW$_{\rm th}$ power station of Angra dos Reis, 
is going to  focus on this high-luminosity approach to provide a full energy spectrum measurement of the oscillation signature. 
The far detector site could be located at 1.5~km from the primary reactor core, under 700~m of granite (1,700~mwe). 
The near detector site could be 300~m distant from the core, and covered by 100~m of granite (250~mwe). 
The detector would be a 500 ton fiducial volume of Gd loaded liquid scintillator. Control of the backgrounds will be crucial. 
If a luminosity of greater than 6,000 GW-ton-year could be achieved, the expected sensitivity would be $\sin^{2}(2\theta_{13})<0.007$.

\end{document}